\documentclass[letter]{ptptex}

\usepackage{graphicx}
\usepackage{bm}
\usepackage{multirow}
\usepackage{subfigure}

\newcommand{\Slash}[1]{\ooalign{\hfil/\hfil\crcr$#1$}}
\newcommand{\eqn}[1]{\label{eq:#1}}
\newcommand{\refeq}[1]{(\ref{eq:#1})}

\newcommand{\Eqs}[2]{Eqs.~(\ref{eq:#1}) and (\ref{eq:#2})}
\newcommand{\Eq}{Eq.~\refeq}
\newcommand{\vecbox}[1]{\mbox{\boldmath{$#1$}}}
\def\leff{${\cal L}$$_{eff}$}

\def\piless{EFT($\Slash{\pi}$)}

\def\veft{$V_{EFT}$}
\def\vph{$V_{ph}$}

\def\1s0{{}^1S_0}
\def\ts1{{}^3S_1}
\def\td1{{}^3D_1}
\def\ltap{\ \raise.3ex\hbox{$<$\kern-.75em\lower1ex\hbox{$\sim$}}\ }
\def\gtap{\ \raise.3ex\hbox{$>$\kern-.75em\lower1ex\hbox{$\sim$}}\ }
\def\simeq{\ \raise.3ex\hbox{$\sim$\kern-.75em\lower1ex\hbox{$-$}}\ }
\def\ket#1{\vert#1\rangle}
\def\bra#1{\langle#1\vert}
\def\braket#1#2{\langle#1\vert#2\rangle}

\notypesetlogo                       

\title{
Explanation of the More Effective Effective Field Theory
}

\author{
Satoshi X. \textsc{Nakamura}\footnote{E-mail:
nakkan@rcnp.osaka-u.ac.jp}%
}

\inst{
Theory Division,
Research Center for Nuclear Physics,
Osaka University\\
Ibaraki, 567-0047, Japan
}

\recdate{November 30, 2004}

\abst{
We show that the so-called {\it more effective} effective field theory (MEEFT)
is essentially equivalent to the proper nuclear effective field theory (NEFT)
in describing low-energy electroweak processes in nuclei.
A key to understanding their equivalence is the relation between a NEFT-based
$NN$-potential and a phenomenological potential through the Wilsonian
renormalization group equation.
We also discuss an important advantage of MEEFT.
}

\begin{document}

\maketitle

%

In the early 1990s, the idea of an effective field 
theory (chiral perturbation theory) was introduced 
into nuclear physics.\cite{weinberg}
In this approach to nuclear physics, to which we will refer as nuclear
effective field theory (NEFT), one derives a $NN$-potential (\veft) and
nuclear current operators from an effective Lagrangian following a
counting rule.
With these nuclear operators, one can describe low-energy phenomena
in nuclear systems, such as
$NN$-scattering and the nuclear response to an electroweak probe.\footnote{
We consider NEFT in the form introduced by
Weinberg\cite{weinberg} in this work.
In another form of NEFT,\cite{pds} on the other hand,
the nuclear potential and nuclear current operator are not explicitly
derived from an effective Lagrangian.
We do not discuss this form of NEFT in the present work.
}
Several authors have constructed \veft's and have shown
the usefulness of NEFT in describing $NN$-scattering.\footnote{
Strictly speaking, the previous NEFT-based constructions of the
$NN$-potential are {\it not} rigorously based on the basic idea of NEFT.
This is pointed out in our recent work.\cite{mine}
}
With regard to electroweak processes in few-nucleon systems, however,
there have been few studies employing a
nuclear potential and nuclear current operator both
obtained from the Lagrangian.\footnote{
There have been many studies of electroweak processes in two-nucleon
systems based on a pionless effective field theory [\piless].
In these studies, KSW-counting\cite{pds}, which is different from
Weinberg's counting, was adopted.
}
Instead, what has been used to this time is the so-called
{\it more effective} effective field theory (MEEFT) developed by Park
et al.\cite{meeft}\footnote{
MEEFT is also referred to as EFT$^*$ in some references.
}
This framework has been applied to 
low-energy electroweak processes in few-nucleon systems, such as
$pp\rightarrow de^+\nu_e$\cite{eft-pp-hep},
${}^3{\rm He}\  p \rightarrow {}^4{\rm He}\  e^+ \nu_e$\cite{eft-pp-hep},
$\nu_e d \rightarrow e^- pp$\cite{eft-nud} and
$\nu d \rightarrow \nu pn$\cite{eft-nud}.

We now briefly explain the calculation of nuclear matrix elements relevant to
these reactions within MEEFT.
We start with an effective Lagrangian, \leff.
When we consider the nucleon and pion as explicit degrees of
freedom, we use a heavy-baryon chiral effective Lagrangian,
as demonstrated, e.g., by Eqs. (4)--(11) of
Ref.~\citen{eft-pp-hep}.
Then, we derive the nuclear current operators following Weinberg's
counting rule and multiply them by a cutoff function, thereby omitting
the nucleonic momentum states above a certain cutoff, $\Lambda$.\footnote{
Because the detailed shape of the cutoff function is not important in
NEFT, we adopt a sharp cutoff in the following.
}
These operators are thus defined in the model space with the
cutoff $\Lambda$.
However, in this calculation, we do not use a
$NN$-potential based on \leff\ but,
instead, a high-precision phenomenological potential, \vph.
We solve the Schr\"odinger equation with such a \vph\
to obtain the nuclear wave functions.
By sandwiching the nuclear operators based on \leff\ between the nuclear
wave functions based on \vph, we obtain the
nuclear matrix elements.
Although this calculational procedure can be carried out in momentum space,
most MEEFT calculations have been done in coordinate space. 
In such a case, there is the additional procedure of
transforming the nuclear current operators obtained 
in the procedure described above
into those expressed in the coordinate space representation.
The nuclear matrix elements are obtained by sandwiching the transformed
operators between the nuclear wave functions obtained in the coordinate
space calculation of the Schr\"odinger equation.

The essence of MEEFT is to use \vph\ instead of \veft.
The motivation for developing MEEFT was to
obtain reaction rates as accurately as possible
within the framework of NEFT.
For that purpose, a realistic nuclear wave function is necessary.
Unfortunately, \veft\ was not as accurate as \vph\ in
describing low-energy $NN$ data.\footnote{
See, however the recent calculation of \veft\ by Epelbaum et
al.\cite{epel} and Entem et al.\cite{entem}
}
Therefore, the MEEFT explained above was adopted.
Clearly, MEEFT is different from the standard
NEFT-based calculational procedure, in which all nuclear operators are
derived from a given effective Lagrangian.
Although MEEFT has met with some criticism,
this hybrid procedure has been phenomenologically
successful, as supported by numerical work.
Given this situation, 
it is very desirable to find a formal basis for MEEFT.\footnote{
We note that Kubodera has given a good argument for the foundation
of MEEFT\cite{kubodera}.
}
The purpose of this paper is to
give a kind of formal foundation to MEEFT
by employing our findings in a recent work.\cite{mine}
Actually, we show that MEEFT is designed to give essentially 
the same results as proper NEFT.

In order to obtain physical quantities for an electroweak process in
a nucleus, we need the nuclear matrix element for the transition.
For simplicity, we consider a two-nucleon system in the following.
We write a nuclear matrix element evaluated with the proper NEFT in the
momentum space as
\begin{eqnarray}
 \eqn{me_eft}
 \bra{\Psi^\prime_{EFT}}\, {\cal O}_{EFT}\, \ket{\Psi_{EFT}} &=&
  \int^\Lambda_0\!\! {d^3k\over(2\pi)^3}\!\! \int^\Lambda_0 \!\!
{d^3k^\prime\over(2\pi)^3}\,
  \braket{\Psi^\prime_{EFT}}{\vecbox{k}^\prime}\bra{\vecbox{k}^\prime}
  \, {\cal O}_{EFT}\, 
\ket{\vecbox{k}}\braket{\vecbox{k}}{\Psi_{EFT}}\ ,
\end{eqnarray}
where ${\cal O}_{EFT}$ is the nuclear current operator derived
from \leff, and $\Psi_{EFT}$ is the nuclear wave function obtained by solving
the Schr\"odinger equation with \veft.
Here, $\ket{\bm{k}}$ represents the free two-nucleon state with
relative momentum $\bm{k}$.
The model space for the nucleon state is specified by
the cutoff $\Lambda$, which
is the maximum allowed magnitude of the relative momentum.
Now, let us compare the above with
the nuclear matrix element evaluated in MEEFT,
\begin{eqnarray}
 \eqn{me_ph}
\bra{\Psi^\prime_{ph}}\eta\, {\cal O}_{EFT}\,\eta \ket{\Psi_{ph}}&=&
  \int^\Lambda_0\!\! {d^3k\over(2\pi)^3}\!\! \int^\Lambda_0 \!\!
{d^3k^\prime\over(2\pi)^3}\,
  \braket{\Psi^\prime_{ph}}{\vecbox{k}^\prime}\bra{\vecbox{k}^\prime}
  \, {\cal O}_{EFT}\, 
\ket{\vecbox{k}}\braket{\vecbox{k}}{\Psi_{ph}}\ ,
\end{eqnarray}
with the projection operator $\eta$ defined by
\begin{eqnarray}
 \eqn{eta}
\eta &=& \int {d^3 q\over (2\pi)^3} \left| \vecbox{q}\, \right\rangle \left\langle \vecbox{q}\,
  \right| \,\, ,  \quad \quad \quad
\left| \vecbox{q} \, \right| \leq \Lambda\ .
\end{eqnarray}
In \Eq{me_ph}, $\Psi_{ph}$ is the nuclear wave function corresponding to \vph.
Although the model space for ${\cal O}_{EFT}$ is smaller than that for
$\Psi_{ph}$, the momentum components beyond $\Lambda$ involved in 
the latter have simply been cut off.
In the following, we demonstrate the
(approximate) equality of the
matrix elements given in \Eqs{me_eft}{me_ph}.
This equality implies the equivalence of
the proper NEFT and MEEFT in describing electroweak processes.

For this purpose, we first recall 
our recent work\cite{mine}, in which the
relation between \veft\ and \vph\ is studied.
We proposed that \veft\ is a parameterization of the model-space
potential ($V_M$) obtained from \vph\ by reducing the model space of \vph\
to an appropriate size for \veft;
the Wilsonian renormalization group (WRG) equation\cite{wilson}
controls the evolution of
the potential due to the model-space reduction.
This point of view concerning the relation between \veft\ and \vph,
which we employ in the following, is the key to
understanding the equivalence of the proper NEFT and MEEFT.

\begin{figure}[t]
\includegraphics[width=65mm,clip]{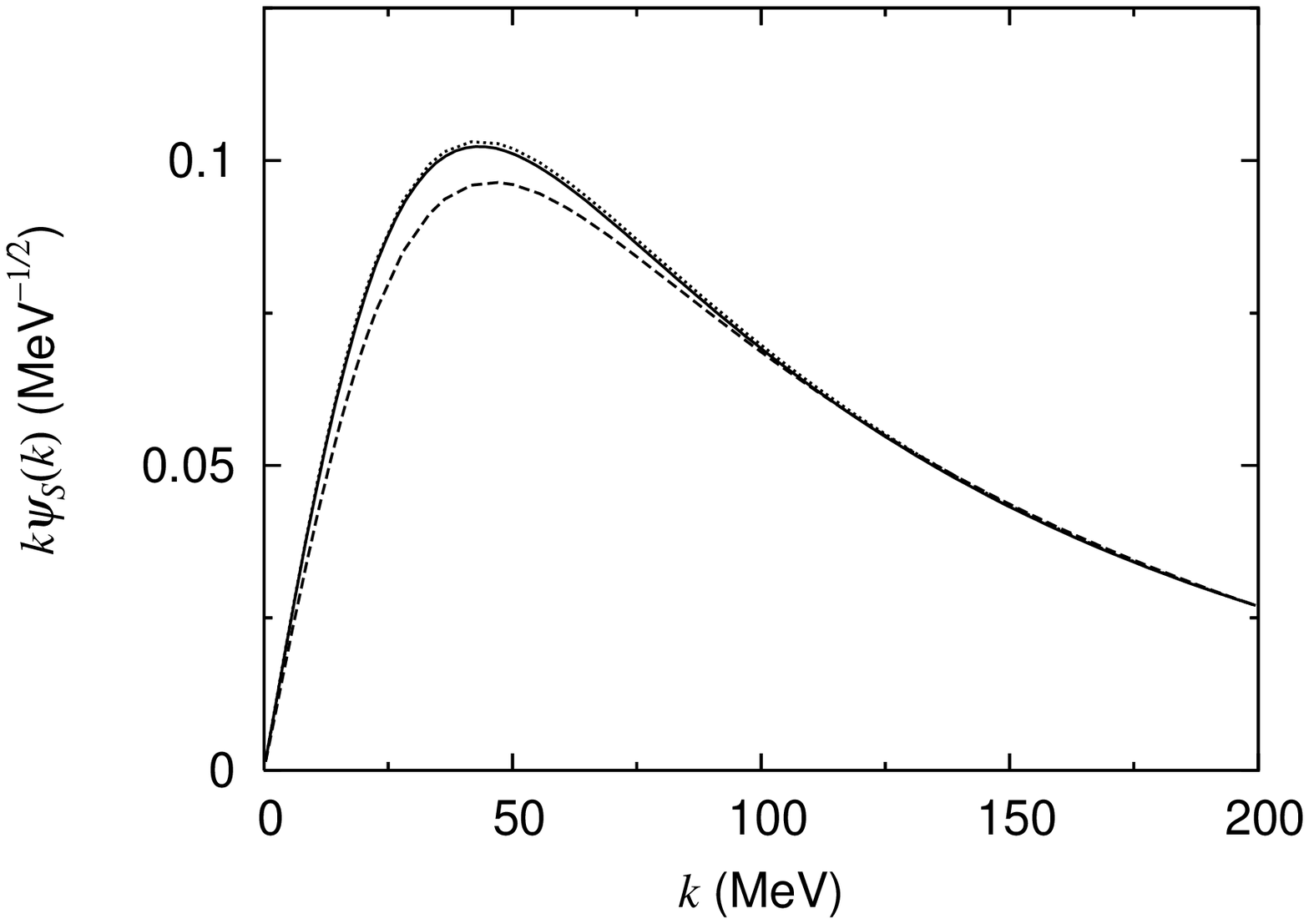}
\hspace{5mm}
\includegraphics[width=65mm,clip]{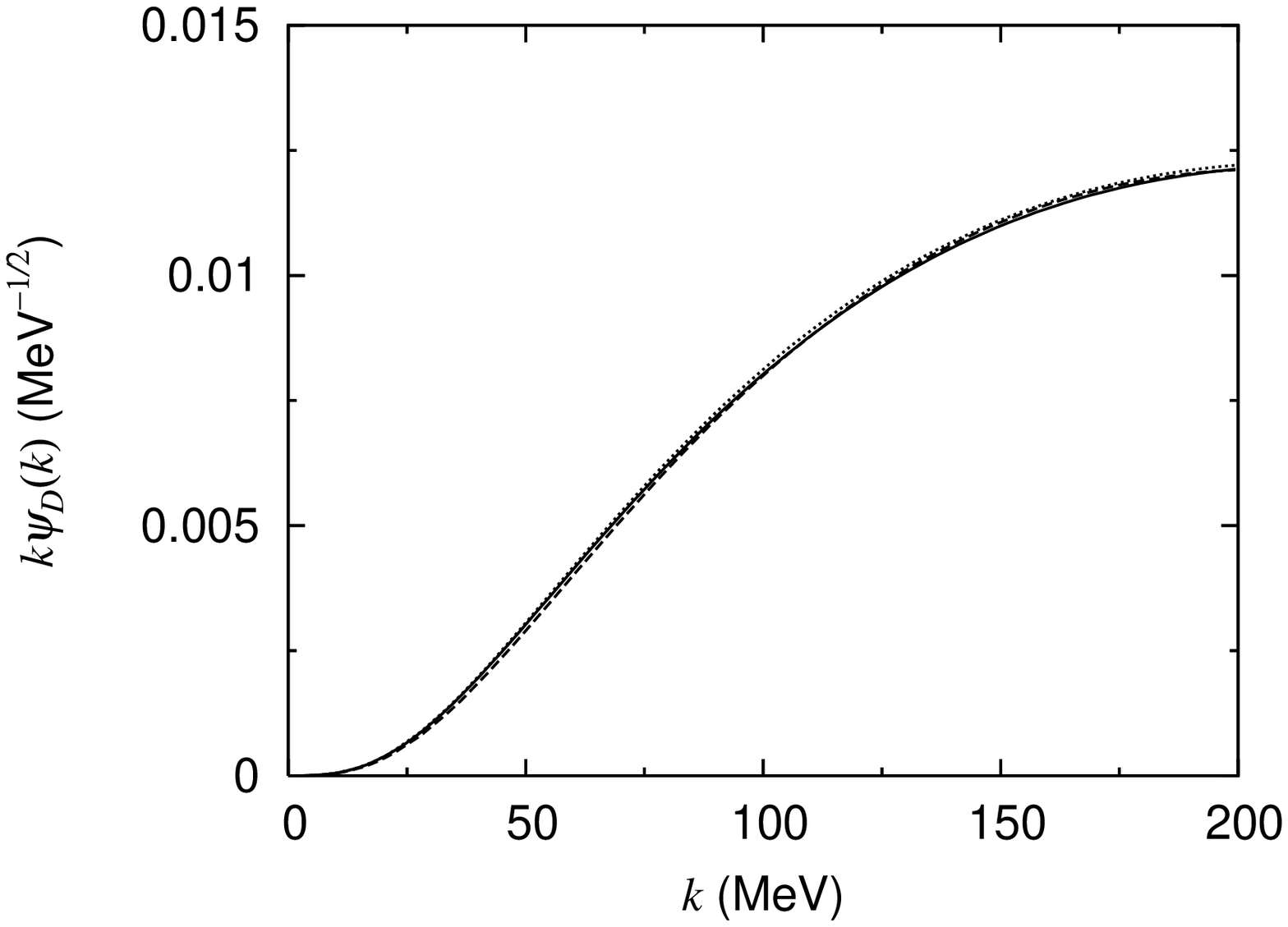}
 \caption{\label{fig_deu_L200}
The radial part of the deuteron wave functions in momentum space. 
The left (right) figure displays the $S$-($D$-)wave.
The solid curve represents $\ket{\Psi_M}$ based on $V_M$
($\Lambda$ = 200 MeV),
while the other curves represent $\ket{\Psi_{EFT}}$ due to \veft.
The dashed curve is from \veft,
including the one-pion-exchange potential, plus contact interactions
 with zero and two derivatives, while
the dotted curve is due to \veft, which
additionally contains terms with four derivatives.
 }
\end{figure}

The WRG equation is given by
\begin{eqnarray}
\eqn{rge}
 {\partial V^{(\alpha)}(k',k;p,\Lambda) \over\partial\Lambda}
= {M\over 2\pi^2} V^{(\alpha)}(k',\Lambda;p,\Lambda){\Lambda^2\over\Lambda^2-p^2}
V^{(\alpha)}(\Lambda,k;p,\Lambda)\ ,
\end{eqnarray}
where $V^{(\alpha)}$ is the $NN$-potential for a given channel (partial wave)
$\alpha$, and $M$ denotes the nucleon mass.
In $V^{(\alpha)}(k',k;p,\Lambda)$,
$\Lambda$ is the cutoff for the relative momentum,
$p$ is the on-shell relative momentum
($p\equiv\sqrt{ME}$, with $E$ being
the kinetic energy of the two nucleons),
and $k$ ($k'$) is the relative momentum before (after) the interaction.
The WRG equation is derived under the condition that the full off-shell
T-matrix elements are invariant with respect to changing $\Lambda$.
This condition for deriving the WRG equation leads to the equation
$\ket{\Psi_{M}}= \eta\ket{\Psi_{ph}}$, where
$\Psi_{M}$ is the wave function corresponding to $V_M$.
Because $V_M$ is {\it to be} parameterized by \veft,
the relation between $\Psi_{EFT}$ and
$\Psi_{ph}$ is given by 
\begin{eqnarray}
 \eqn{psi_eft_ph}
 \ket{\Psi_{EFT}} \simeq \ket{\Psi_{M}}= \eta\ket{\Psi_{ph}}\ .
\end{eqnarray}
Thus, we immediately obtain the relation
\begin{eqnarray}
 \eqn{me_ph3}
\bra{\Psi^\prime_{EFT}}\, {\cal O}_{EFT}\, \ket{\Psi_{EFT}}
&\simeq&
 \bra{\Psi^\prime_{ph}}\eta\, {\cal O}_{EFT}\,\eta \ket{\Psi_{ph}}\ .
\end{eqnarray}
This equation expresses the approximate equality of
a nuclear transition matrix element evaluated
with the proper NEFT (l.h.s.) and that obtained with MEEFT (r.h.s.).
We now discuss the accuracy of the approximate equalities in
\Eqs{psi_eft_ph}{me_ph3}.
First, note that \veft\ is a parameterization of $V_M$.
Therefore, if one includes, in principle,  
terms of sufficiently high order in \veft,
one obtains a \veft\ that is essentially the same as $V_M$.
Actually, we have shown in Ref.~\citen{mine}
that the NEFT-based parameterization of $V_M$
is very effective
even for low-order perturbations, which are practically feasible.
We also showed that \veft\ yields values of the phase shifts
and the deuteron binding energy that are quite close to those given by $V_M$.
Furthermore, we confirmed that the deuteron wave function generated by
$V_M$ is accurately approximated by that obtained with \veft.
Here, we simply compare $\ket{\Psi_M}$ with $\ket{\Psi_{EFT}}$ for the
deuteron state in Fig. \ref{fig_deu_L200}.
It is seen there that the approximate equality expressed in \Eq{psi_eft_ph}
is indeed quite good.
Therefore, we conclude that
\Eqs{psi_eft_ph}{me_ph3} represent very good approximations.
We have thus shown that
the proper NEFT and MEEFT are essentially equivalent in describing
electroweak processes.

Even though this equivalence can be seen clearly when we employ the
momentum space representation, as shown above, once the equivalence is
established, the choice of the representation is a matter of taste,
as we can also employ the coordinate space representation.
A nuclear matrix element evaluated with MEEFT in the momentum space 
[\Eq{me_ph}] can be rewritten as
\begin{eqnarray}
 \eqn{me_mom}
 \bra{\Psi^\prime_{ph}}\eta\, {\cal O}_{EFT}\, \eta \ket{\Psi_{ph}} &=&
  \int^\infty_0 {d^3k\over(2\pi)^3}\int^\infty_0 {d^3k^\prime\over(2\pi)^3}
  \braket{\Psi^\prime_{ph}}{\vecbox{k}^\prime}\bra{\vecbox{k}^\prime}
  \eta\, {\cal O}_{EFT}\, \eta
\ket{\vecbox{k}}\braket{\vecbox{k}}{\Psi_{ph}}\ .
\end{eqnarray}
Because of the presence of the projection operator $\eta$,
we are able to safely extend the domain of integration to infinity.
This is just an ordinary matrix element evaluated in the momentum space,
and therefore it is straightforward to express
$\bra{\Psi^\prime_{ph}}\eta\, {\cal O}_{EFT}\,\eta \ket{\Psi_{ph}}$
using the coordinate space representation 
through the Fourier transformation.
The operator $\eta\,{\cal O}_{EFT}\, \eta$ in the coordinate space is
generally non-local.
We have thus shown that the MEEFT calculation in the coordinate space is
(approximately) equivalent to the proper NEFT calculation.

Strictly speaking, however, MEEFT calculations carried out previously
are not the
same as the coordinate space calculation explained above.
However, this is due only to the use of a cutoff function in those
MEEFT calculations.
Allow us to explain this point further.
In the above case,
$\bra{\bm{k}^\prime}\eta\, {\cal O}_{EFT}\, \eta\ket{\bm{k}}$ is zero in
the cases $|\vecbox{k}|>\Lambda$ and/or
$|\vecbox{k}^\prime|>\Lambda$.
In the previous MEEFT calculations, by contrast, a cutoff function on
the momentum transfer,
$|\vecbox{k}^\prime-\vecbox{k}|$, was used for
convenience.
However, this choice of the cutoff function is not consistent with the proper
way of integrating out the high-momentum states in NEFT.
With this cutoff, the model-dependent, high-momentum
components of the wave function enter into the nuclear matrix elements.
Note, however, that despite this fact, the treatment of low-energy phenomena
(low-momentum components) is the same as that in the proper NEFT.
The difference between the two treatments of
short-distance phenomena is expected to be largely eliminated by shifting the
couplings of the contact interactions involved in ${\cal O}_{EFT}$, 
because the {\it details} of short-distance physics are not important for
describing low-energy reactions.
This conjecture is supported by the phenomenological
success of MEEFT.

Finally, we discuss the advantages of MEEFT.
The first advantage is that we can use a high precision phenomenological
$NN$-potential. As stated above, this was the motivation of developing
MEEFT,
and let us escape from
constructing \veft\
as accurate as \vph.
The second advantage, which
is probably more important, is as follows.
We consider an electroweak process
in which the initial and/or final state is the
deuteron state.
In this case, we need low-momentum components of the {\it normalized}
deuteron wave function in the model space.
As discussed in Ref.~\citen{mine}, however, this normalization requires the
high-momentum components of the wave function,
or equivalently, information about the {\it details} of short-distance
physics.
Such information is not
available if one employs the model-space framework of NEFT.
Therefore, in that case, it is necessary to rely on a model of the
short-distance
physics for the normalization.
Given this situation,
MEEFT is a useful procedure to incorporate such a model into a
NEFT-based calculation, because with it,
the normalization given by the high precision \vph\ can be used.

\section*{Acknowledgements}
I am very grateful to Prof. H. Toki for
his critical reading of the draft and valuable discussions on this work.
I also acknowledge Dr. S. Ando for useful discussions.


\begin{thebibliography}{99}

\bibitem{weinberg}
S. Weinberg, Phys. Lett. B \andvol{251,1990,288}; Nucl. Phys. B \andvol{363,1991,3}.

\bibitem{pds}
D. B. Kaplan, M. J. Savage and M. B. Wise, Phys. Lett. B
	\andvol{424,1998,390}; Nucl. Phys. B \andvol{534,1998,329}.

\bibitem{mine}
S. X. Nakamura, Prog. Theor. Phys. \andvol{114,2005,77}.

\bibitem{meeft}
T.-S. Park, K. Kubodera, D.-P. Min and M. Rho, Nucl. Phys. A \andvol{684,2001,101}.

\bibitem{eft-pp-hep}
T.-S. Park, L. E. Marcucci, R. Schiavilla, M. Viviani, A. Kievsky, S. Rosati, K. Kubodera, D.-P. Min and M. Rho, Phys. Rev. C \andvol{67,2003,055206}.

\bibitem{eft-nud}
S. Ando, Y. H. Song, T.-S. Park, H. W. Fearing and K. Kubodera, Phys. Lett. B \andvol{555,2003,49}.

\bibitem{epel}
E. Epelbaum, W. Gl\"ockle,  and Ulf-G. Mei{\ss}ner,
Nucl. Phys. A \andvol{747,2005,362}.

\bibitem{entem}
D. R. Entem and R. Machleidt,
Phys. Rev. C \andvol{68,2003,041001}.

\bibitem{kubodera}
K. Kubodera, nucl-th/0308055.

\bibitem{wilson}
K. G. Wilson and J. B. Kogut, Phys. Rep. \andvol{12,1974,75}.\\
J. Polchinski, Nucl. Phys. B \andvol{231,1984,269}.\\
M. C. Birse, J. A. McGovern and K. G. Richardson, Phys. Lett. B \andvol{464,1999,169}.

\end{thebibliography}
\end{document}